# Measurement of spin waves and activation volumes in superparamagnetic Fe films on GaAs(100)

R. L. Stamps[1], A. Stollo[2], M.Madami[2], S. Tacchi[2], G. Carlotti [2,3], G. Gubbiotti[4], M. Fabrizioli[5], J. Fujii[5]

[1] School of Physics, University of Western Australia, 35 Stirling Highway, Crawley WA 6009, Australia

[2] INFM, Dipartimento di Fisica, Universiti di Perugia, Via Pascoli, 06123 Perugia, Italy

[3] INFM-CNR Research Center S3, Via Campi 213/a, 41100, Modena (Italy)

[4] Research center  SOFT-INFM-CNR, Universita' di Roma "La Sapienza", I-00185, Roma, Italy

[5] INFM, TASC Laboratory, Area Science Park, S.S.14, Km 163.5, I-34012 Trieste, Italy

**Abstract**

Spin wave frequencies are observed in ultra-thin Fe/GaAs(100) films at temperatures where the spontaneous zero field magnetization is zero. The films exhibit good cyrstalline structure, and the effect of magnetic anisotropies is apparent even though no zero field spin wave energy gap exists. An analysis is given in terms of a superparamagnetic model  in which the film is treated as a network of non-interacting single domain magnetic islands. A spin wave analysis provides a means to separate measured values of anisotropy parameters from products involving anisotropy and island volume. In this way, a measure of the activation volume associated with superparamagnetic islands is obtained for different Fe film thicknesses.  Results suggest that the island lateral area increases with increasing film thickness.

75.70.Cn, 75.60.-d,75.50.Lk

# I. INTRODUCTION

The ability to grow epitaxially ferromagnetic films with only a few atomic layers is a great experimental achievement [1]. These films provide useful model systems in which basic aspects of magnetic ordering in low dimensions can be studied and theories of magnetism can be tested [1,2]. Self organized ferromagnetic metal structures can emerge during the first few atomic layers of growth on semiconductor substrates, with volumes and shapes that allow single domain ferromagnetic particle behavior. Crystalline structure and orientation during island formation can be controlled to a large extent by choosing the substrate and growth conditions appropriately. Continuous films appear when islands merge, and one expects some sort of transition from superparamagnetic behaviour to ferromagnetic behavior across a range of film thicknesses.

Experimentally it is not a simple matter to determine when islands merge in terms of ferromagnetic properties. Structural studies of dense island structures cannot determine how strongly magnetic islands may interact. A relevant question is therefore as follows. When the average thickness during film growth is increased, does the *magnetic* behavior evolve into that of a three dimensional film from that of independent superparamagnetic particles or instead from that of a continuous quasi-two dimensional magnetic film? In both cases a transition occurs that can be identified with some temperature: in the former case the transition occurs at a 'blocking' temperature, whereas in the latter case it occurs at a critical temperature for long range order. Also in both cases, anisotropy plays a deciding role although how magnetic anisotropies evolve with particle size and film thickness is not well understood [3,4].

We present results from inelastic light scattering experiments on ultrathin films of Fe grown on GaAs which self organize into densely packed islands. In a previous work it has been shown that this system is superparamagnetic for a range of Fe film thicknesses between three and four monolayers [5]. Steinmuller, et al., showed that it is possible to observe spin excitations from paramagnetic Fe films in this thickness range [6]. Here, we examine effects of magnetic anisotropies on resonances measured in the paramagnetic region, and show for the first time that one can extract separately values for the magnetic anisotropy and activation volumes. Our results are consistent with a picture of film formation by merging of islands during deposition as coverage increases.

# II. THEORETICAL BACKGROUND

The blocking temperature $T_B$ of superparamagnetism describes thermal instability of regions which behave as effectively independent, single domain magnetic entities. Consequently there are requirements for length and time scales in order for superparamagnetism to exist. Single domain reversal is possible when the energy involved in reversing a ferromagnetic region is small compared to that of forming domain boundary walls. This is possible in nanoparticles since it can be energetically unfavorable to retain a domain wall within a structure with dimensions less than or equal to a domain wall width.

Superparamagnetism involves thermally driven, irreversible reorientations of the angular momentum associated with the magnetic moment. The minimum time involved is determined by dissipation of energy during precession of the magnetic moment. Typical time scales are on the order of 10 nanoseconds, determined in metallic ferromagnets largely through coupling with the conduction spin system facilitated by spin orbit coupling to the lattice. The thermal reversal rate $1/\tau$ follows a chemical rate type law of the form $1/\tau = f_o \exp\left(-\frac{V\varepsilon_a}{k_B T}\right)$ where $V\varepsilon_a$ is the energy associated with reversing some volume $V$ with activation energy density $\varepsilon_a$. The prefactor $f_o$ is an 'attempt' frequency determined by details of exactly how the magnetic moment fluctuates about the equilibrium and unstable configurations.

The time scale for superparamagnetism is much smaller than the acquisition time for quasi-static measurements of magnetization such as conventional magnetometry. The time scales question is relevant for high frequency probes such as inelastic light scattering since the low energy spin wave modes and resonances occur in the low GHz regime for ferromagnetic metals. A resonant response can be measured at temperatures above $T_B$ as long as the magnetization of an ensemble of particles varies slowly with respect to the precession frequency of spin waves. This is possible for transition metal particles with high anisotropy and magnetic moment.

Low frequency spin waves have associated wavelengths much longer than the nanometer dimensions characteristic of single domain particles, and thereby sample an average over an ensemble of magnetic islands. Above $T_B$, the spin excitation is therefore produced by fluctuations about a configurational average of many islands. Within each particle, the precession frequency is determined by effective anisotropy and magnetostatic fields, and a long wavelength spin wave measures an ensemble average over possible response frequencies.

An interesting case arises when the anisotropy axes are aligned uniformly throughout the ensemble. This results in a common easy and hard direction for all particles. For temperatures above $T_B$, the magnetization of the ensemble is described by an anisotropic paramagnetic susceptibility. For a magnetic field $H$ applied along the easy direction, the induced magnetization in the high temperature limit is $M=\chi_\parallel H$. When applied along the hard direction, one has instead $M=\chi_\perp H$.

The long wavelength spin wave for an ensemble of non-interacting nanoparticles is essentially the ferromagnetic resonance frequency with small magnetostatic corrections at finite wavelengths and film thicknesses. These effects can be neglected to good approximation for the film thicknesses and wavelengths studied in this experiment. The ferromagnetic resonance frequency for a planar geometry with the y-axis perpendicular to the sample surface and with the field in the in-plane easy direction then obeys the Kittel resonance condition [7]:

$$\omega_{easy}^2 = \gamma^2 \left[ H + H_a + 4\pi M (N_y - N_x) \right] \left[ H + H_a + 4\pi M (N_z - N_x) \right],$$

(1)

where the anisotropy field is $H_a$, $M$ is the thermally averaged magnetization along the direction of the applied field, and the $N_i$ are demagnetization factors.

The anisotropy field is a function of $M$, and therefore equal to $(2K/M_S^2) \chi_\parallel H$ where $M_S$ is the zero temperature saturation magnetization. In the high temperature regime, the frequency is linear in field with a slope that depends on anisotropy energy density $K$ and the static susceptibility $\chi_\parallel$. A similar argument applies for the ferromagnetic resonance with the field in the hard in plane direction,

$$\omega_{hard}^2 = \gamma^2 \left[ H + 4\pi M (N_y - N_z) \right] \left[ H - H_a - 4\pi M (N_z - N_x) \right]$$

(2)

in which case $\omega \propto H$ with a constant of proportionality that depends on $K$ and $\chi_\perp$.

Measurement of the long wavelength spin wave frequencies as a function of applied field therefore gives information about the magnetic anisotropy *and* activation volume through the static susceptibilities. A key point is that the activation volume enters only in the static susceptibility. It can be shown that for a single particle of volume $V$ in the high temperature regime, $\chi_\parallel = (g\mu_B)^2 \beta (1/3 + 4\beta KV/45)$ and $\chi_\perp = (g\mu_B)^2 \beta (1/3 - 4\beta KV/45)$ where $\beta = 1/(k_B T)$, and $K$ is the anisotropy energy density. The easy and hard direction susceptibilities differ by the sign on the activation volume term. This feature can be used to extract a measure of $V$ separately from $K$.

The analysis sketched out above was used to interpret inelastic light scattering data taken from Fe layers deposited on GaAs substrates.

### III. EXPERIMENTAL DETAILS

The Fe/GaAs(001) films were prepared at the GHOST laboratory of the Perugia of University [http://ghost.fisica.unipg.it]in a ultra-high-vacuum (UHV) chamber which was specially designed to be easily and rapidly interfaced with our optical table in order to perform *in-situ* BLS measurements. To this purpose, the specimen can be positioned close to an optical viewport, within the poles of an external four inches electromagnet [8]. This chamber is equipped with usual UHV tools such as low-energy electron diffraction (LEED), reflection high-energy electron diffraction (RHEED), and Auger electron spectroscopy (AES). In order to gain a deeper insight into both the morphology and magnetism of the studied films, Fe/GaAs(100) films were also prepared following the same procedure in another chamber, at TASC laboratory in Trieste, where, in addition to the LEED-AES measurements, there is also the possibility to perform *in-situ* magneto-optical kerr effect (MOKE) analysis and Scanning Tunneling Microscopy (STM). Fe films with nominal thickness in the range between 0.7 and 5 Å were grown on commercial Si-doped GaAs(001) wafers, following the same procedure in both chambers. Before deposition the substrate was prepared by several cycles of $Ar^+$-ion sputtering (5 x $10^{-6}$ mbar, 600 eV) each followed by 30 minutes annealing at 900 K. Fe films were then deposited by means of electron beam evaporation at a pressure not exceeding 2.0 x $10^{-9}$ mbar. The evaporation rate, which was typically 1 Å/min, was monitored by means of a quartz microbalance while the substrate was kept at room temperature. LEED-AES analysis was performed on all samples in order to verify the good epitaxial growth of Fe films and their chemical purity.

## IV. RESULTS AND DISCUSSION

From STM analysis it turned out that continuous films appeared only for thicknesses around 5 Å. For smaller thicknesses there is a regular array of nearly-circular islands, whose diameter increases from about 20 to 40 Å as the nominal film thickness goes from 0.7 to 5 Å. This can be clearly seen in Fig. 1(a) where an 850 by 850 Å STM image is shown for an equivalent film thickness of 0.7 Å. Magnetic hysteresis loops were measured as a function of temperature, in the range 150-350 K. The magnetic signal was revealed for nominal film thickness larger than 3.5 Å. A paramagnetic behavior occurs at room temperature for all the films studied, including the 5 Å sample whose loops are shown in Fig. 1(b). It can be seen that for this specimen, a ferromagnet/paramagnet transition occurs between 260 and 230 K. The high frequency magnetic properties of these superparamagnetic films have been investigated in-situ by systematic room-temperature BLS measurements of the spin-wave frequency as a function of the intensity of the applied magnetic field (*H*) between 0 and 4 kOe and the in-plane direction of the applied magnetic field with respect to the [010] direction of the GaAs(001) substrate ($\phi_H$). About 200 mW of monochromatic P-polarized light, from a solid state Laser (532 nm line), was focused onto the sample surface using a camera objective of numerical aperture 2 and focal length 50 mm. The back-scattered light was analyzed by a Sandercock type (3+3)-pass tandem Fabry-Pérot interferometer [9], with typical acquisition time of about 15 minutes. The external dc magnetic field was applied parallel to the surface of the film and perpendicular to the plane of incidence of light.

Typical BLS spectra recorded applying the external field along either the easy or the hard direction are shown in Fig. 2 for the 4 Å thick Fe film. Resonance peaks are clearly visible both (a) above and (b) below the blocking temperature.

The frequencies of magnetic peaks identified in the light scattering spectra are plotted in Fig. 3 as functions of the applied field for three different Fe thicknesses: (a) 3.5 Å, (b) 4 Å, and (c) 4.5 Å. Results are shown for the field applied along an easy (squares) and hard (circles) direction. For all film thicknesses considered, the BLS data were collected at room temperature, i.e. in the paramagnetic region, as found by the MOKE analysis.. The blocking temperature increases as thickness is increased, so that room temperature is not far above $T_B$ for the 4.5 Å film. The frequencies increase linearly with applied field except for the smallest fields. The nonlinearity is most pronounced for the thickest film.

Fits to the measured frequencies are shown by the solid lines in Fig. 3. The frequencies for the ferromagnetic resonance mode of a single domain island has been calculated using temperature dependent static susceptibilities $\chi_\perp$ and $\chi_\parallel$. The susceptibilities are calculated from a single classical vector model for the magnetic moment of an island of volume $V$. The orientation of this vector away from the easy direction is specified by an angle $\theta$. The average projection of the moment along the the easy direction is therefore defined as $\langle \cos\theta \rangle$, and the average is calculated from

$$\langle \cos\theta \rangle = \frac{\int_0^\pi \cos\theta\, e^{-V\varepsilon_a \beta} \sin\theta\, d\theta}{\int_0^\pi e^{-V\varepsilon_a \beta} \sin\theta\, d\theta}.$$

(3)

Cylindrically shaped dots are assumed, and the magnetic moment is assumed to remain in the direction of the field applied in the plane of the substrate. With the field applied along an easy direction, the energy $\varepsilon$ is given by $\varepsilon_a = HM\cos\theta + K\cos^2\theta$. For the field along a hard direction, we calculate $\langle \sin\theta \rangle$ using $\varepsilon_a = HM\sin\theta + K\cos^2\theta$.

Analytic expressions for the susceptibilities can be calculated in the limit of large temperature and defined by $\chi_\perp = \langle \sin\theta \rangle /H$ and $\chi_\parallel = \langle \cos\theta \rangle /H$. The resulting susceptibilities, valid to second order in temperature, were given earlier. For arbitrary temperatures above $T_B$, the average defined in Eq. (1) can be evaluated numerically.

Explicit expressions for the resonant frequencies can be found as outlined above. Fits to the data were made using the following parameters. In all cases, the saturation magnetization $4\pi M_S$=2.2 kOe, $2\pi\gamma$=18 GHz/kOe, and $K$=0.87 erg/cm$^3$. The particle shape factors, corresponding to an oblate ellipsoid with a diameter to thickness ratio of 4, are : $N_x$=$N_z$=0.15 and $N_y$=0.7. Using these values, a consistent set of fits were sought for easy and hard field directions for all films by adjusting only the activation volume $V$. The best fits were obtained using volumes of 20 × 10$^{-19}$ cm$^3$ for the 4.5 Å film, 15 × 10$^{-19}$ cm$^3$ for the 4 Å film, and 9 × 10$^{-19}$ cm$^3$ for the 3.5 Å film.

The activation volume is therefore seen to increase with effective Fe film thickness. Corresponding effects in the field dependence of the frequencies can be observed. At high temperatures, the frequencies are linear in field regardless of field orientation along the easy or hard direction. More complicated field dependencies appear for lower temperatures approaching $T_B$. The fits to the data are consistent with this. All measurements were made at room temperature, but $T_B$ increases as the Fe film thickness is increased. For this reason, pronounced nonlinearity of frequency with respect to field is seen in the plots of Fig. 3 at small fields for the thickest film.

An interesting relation between lateral area and activation volume can be inferred from these results. If we take the film thickness as a nominal measure of the average particle height, then the particle cylinder base area $A$ can be estimated from the ratio $V/t_{Fe}$ where $t_{Fe}$ is the Fe thickness. This area increases with increasing $t_{Fe}$: $A$=4.4 × 10$^{-11}$ cm$^2$ for the 4.5 Å film, $A$=3.8 × 10$^{-11}$ cm$^2$ for the 4 Å film, and $A$=2.6 × 10$^{-11}$ cm$^2$ for the 3.5 Å film. These areas correspond to island diameters on the order of 10 nm which are slightly larger than those obtained by STM. This discrepancy is not surprising, however, because the activation volumes pertain to *magnetic* volumes, and may not be exactly the physical volumes. Furthermore, the increase in $A$ with $t_{Fe}$ suggests that the lateral size of the islands increases as the effective Fe film thickness increases. This is reasonable if continuous film behavior follows from merging of island during film growth.

In summary, we have observed anisotropic susceptibilities through measurement of inelastic scattering of light from magnetic excitations in superparamagnetic films above the blocking temperature. The films are Fe grown epitaxially on GaAs (100). We are able to obtain separate measures for the anisotropy energy density and activation volumes associated with the superparamagnetic behavior by measuring frequencies with the static field applied along easy and hard directions. The lateral sizes of islands appear to increase more rapidly than height with increasing Fe film thicknesses, and eventually merge to behave magnetically as continuous films.

**Acknowledgements**

We thank the CNR and Australian Research Council for support of this work.

(a)

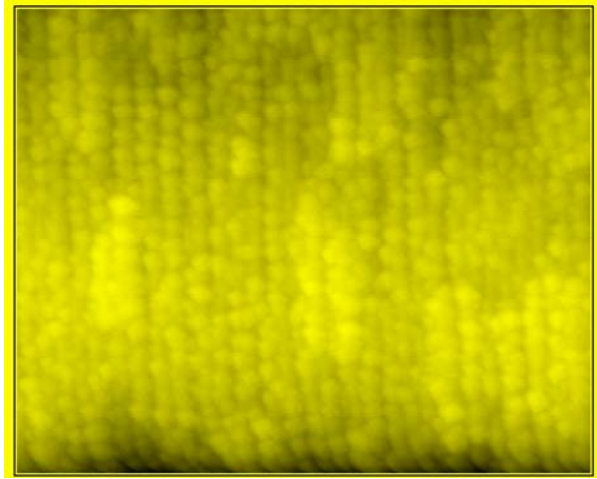

(b)

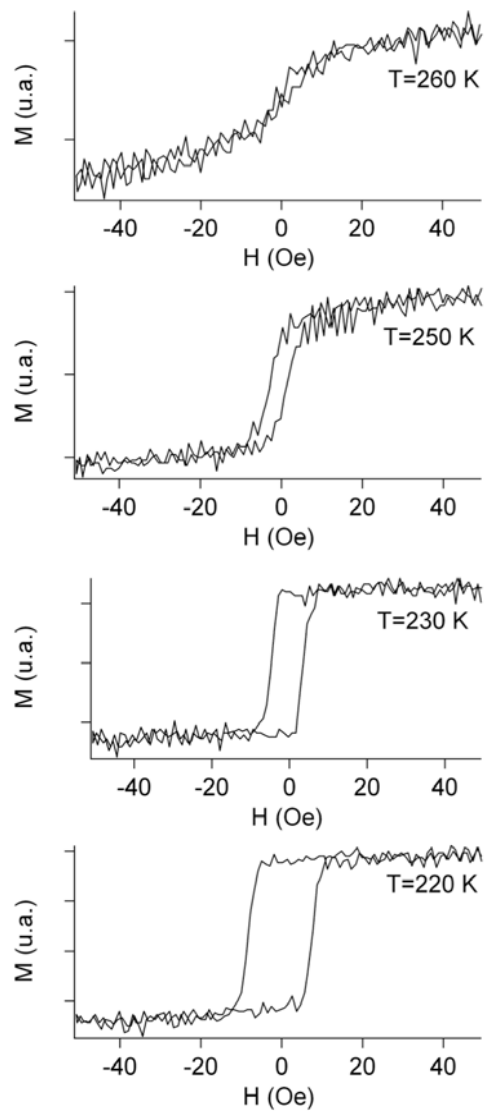

Figure 1. (a) Scanning Tunnelling Microscope image of a 0.7 Å thick Fe film grown on GaAs. (b) Magnetization curves of a 5 Å thick on a Fe/GaAs film measured at different temperatures.

(a)

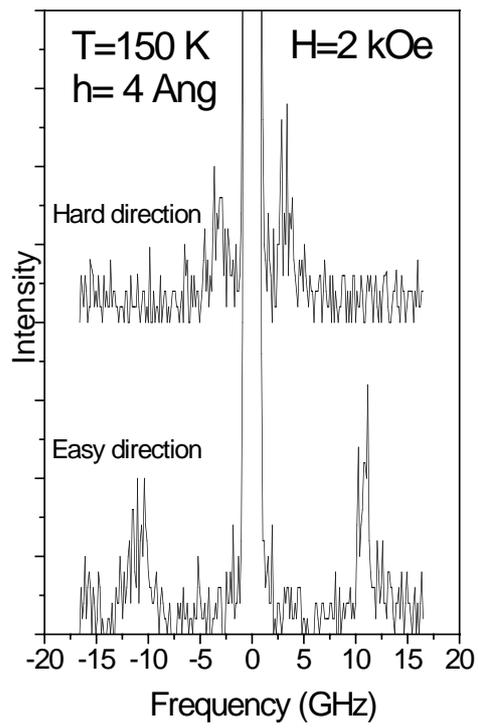

(b)

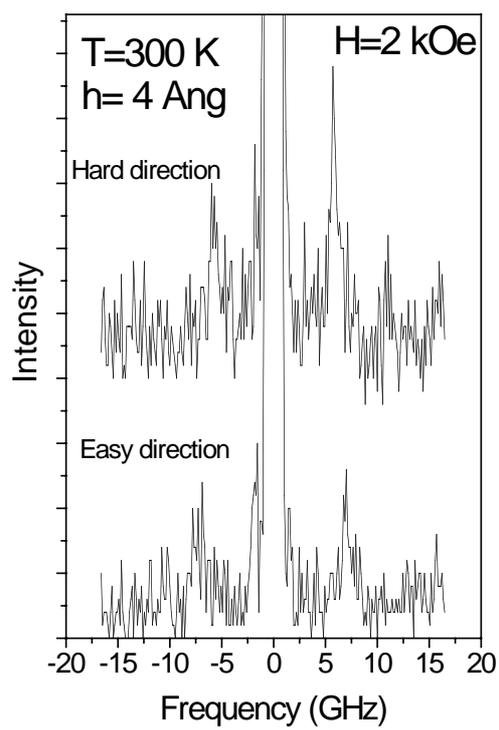

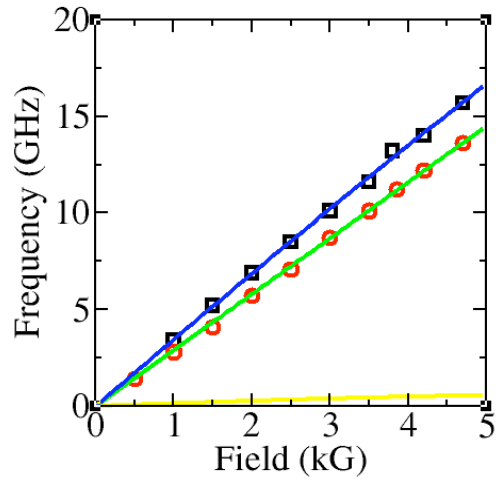

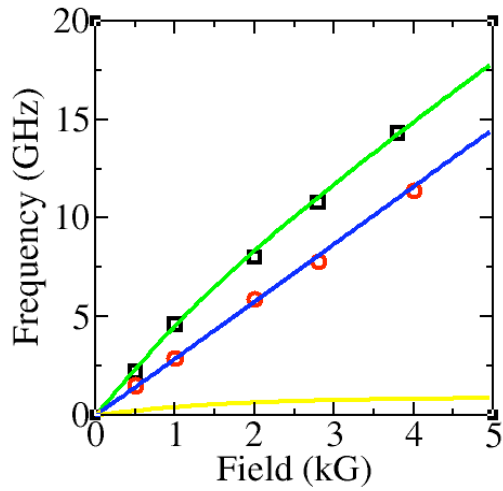

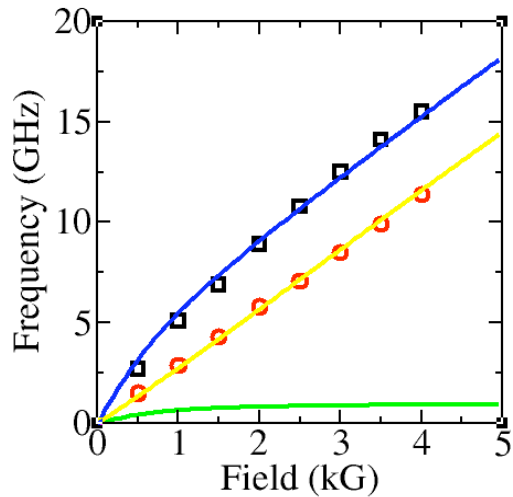

Figure 3. Frequencies determined from Brillouin light scattering as a function of applied field at room temperature. Results for the field applied along an easy direction are shown by squares, and results for the field applied along a hard in plane direction are shown by circles. Fits are indicated by the solid lines through the data points. The calculated induced magnetization is also shown (normalized to unity). Results for three film thicknesses are shown: (a) 3.5 Å, (b) 4 Å, and (c) 4.5 Å.